\documentclass[twocolumn,prb,aps,showpacs,amsmath,amssymb]{revtex4}

\usepackage{graphicx}
\usepackage{dcolumn}
\usepackage{bm}

\renewcommand{\vec}[1]{{\bf#1}}

\begin{document}

\title{Anomalous Cherenkov spin-orbit sound}

\author{Sergey Smirnov}
\affiliation{Institut f\"ur Theoretische Physik, Universit\"at Regensburg,
  D-93040 Regensburg, Germany}

\date{\today}

\begin{abstract}
The Cherenkov effect is a well known phenomenon in the electrodynamics of fast
charged particles passing through transparent media. If the particle is
faster than the light in a given medium, the medium emits a forward light
cone. This beautiful phenomenon has an acoustic counterpart where the role of
photons is played by phonons and the role of the speed of light is played by
the sound velocity. In this case the medium emits a forward sound cone. Here,
we show that in a system with spin-orbit interactions in addition to this {\it
  normal} Cherenkov sound there appears an {\it anomalous} Cherenkov sound
with forward and {\it backward} sound propagation. Furthermore, we demonstrate
that the transition from the normal to anomalous Cherenkov sound happens in a
{\it singular} way at the Cherenkov cone angle. The detection of this acoustic
singularities therefore represents an alternative experimental tool for the
measurement of the spin-orbit coupling strength.
\end{abstract}


\maketitle

Purely electric manipulation of the electron spin is, no doubt, the core idea
of modern spintronics \cite{Zutic,Fabian}.  Systems with spin-orbit
interactions (SOI) represent a distinct platform for the practical
implementation of this idea by way of concrete spintronic devices such as a
spin transistor \cite{Datta}. In two-dimensional semiconductor
heterostructures usually the Bychkov-Rashba SOI (BRSOI)
\cite{Bychkov}, resulting from
the structure inversion asymmetry, and linear Dresselhaus  \cite{Dresselhaus}
SOI (DSOI), resulting from the inversion asymmetry of
crystal structure of the bulk host material, are of most practical importance
in the spin dynamics. The cubic Dresselhaus term is less significant but in
exotic situations, when the coupling strengths of BRSOI and DSOI are equal, it
becomes the main term violating the SU(2) symmetry and thus plays a crucial
role in limiting the electron spin life time as has been recently demonstrated
in the fascinating experiments on persistent spin helix \cite{Koralek,Fabian_1}.

Most works on systems with SOI focus on electron spin and
charge dynamics, in particular, on pure spin currents generation: by means of
intrinsic spin-Hall effect \cite{Murakami,Sinova}, polarized light \cite{Zhou}
or spin ratchets \cite{Smirnov}.

Also fundamental experimental research on spin-orbit coupling (SOC) strength
mainly addresses the electron degrees of freedom, through Shubnikov-de Haas
oscillations \cite{Luo,Schaepers}, photocurrents \cite{Ganichev} or optical
monitoring of the angular dependence of the electron spin precession on the
electron motion direction with respect to the crystal lattice \cite{Meier}.

This trend, which puts the electron degrees of freedom in the center of the
research is clear. From one side it is explained by the fact that exactly
electron dynamics represents the source of SOC. SOI
is an outcome of special relativity where in the
reference frame of a moving electron electric fields transform into magnetic
ones. These magnetic fields interact with the electron spin removing the spin
degeneracy. From the other side it is explained by significant advances in
experimental technique dealing with electrons.

However, in real systems the electron degrees of freedom may interact with
degrees of freedom of different nature. It is therefore challenging to study
the traces of SOC on the particles interacting
with electrons. One possibility is provided by the lattice vibrations of the
heterostructure host crystal. Indeed, the electron orbital degrees of freedom
are electrostatically coupled to the orbital degrees of freedom of the crystal
lattice. Since the electron orbital dynamics in systems with SOI
depends on spin, it is clear that the lattice dynamics will be in
a certain way modified by electron SOI.
\begin{figure}
\includegraphics[width=7.6 cm]{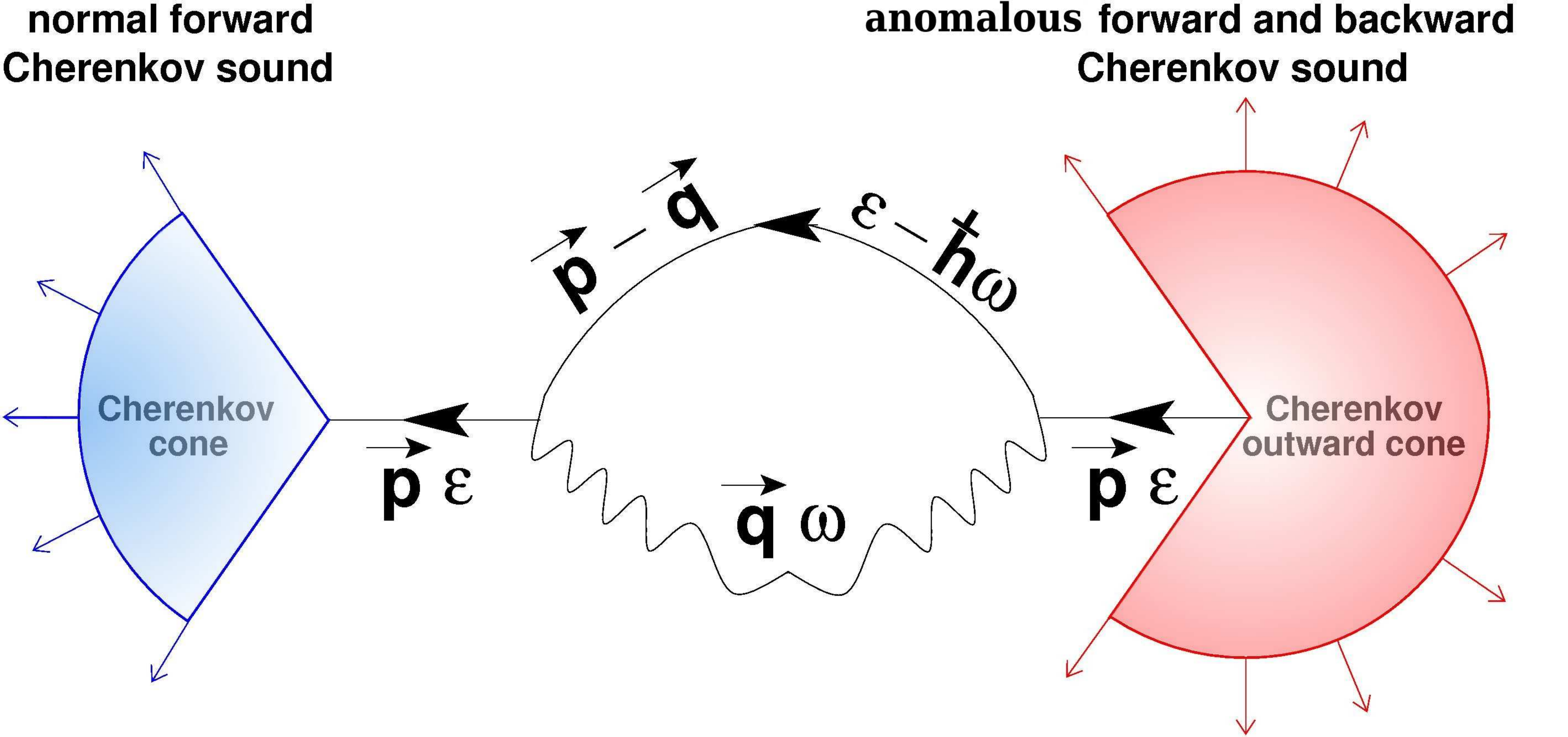}
\caption{\label{figure_1} (Color online) The Feynman diagram of ACE describing
  the process of the acoustic phonons excitation by an incident electron with
  momentum $\vec{p}$ and energy $\varepsilon$. The phonons with momentum
  $\vec{q}$ are emitted in the normal forward sound cone and, for SOC
  systems, in the anomalous forward and backward outward sound cone.}
\end{figure}

The interaction between electrons and the crystal lattice is described in
terms of quantized lattice vibrations, referred to as phonons, and it has been
widely studied in systems with SOI. However, the research
mostly centered again on the impact of phonons on the electron degrees of
freedom, either charge or spin. Examples date back to research on spin
relaxation due to Dyakonov-Perel' \cite{Dyakonov} mechanism involving electron
scattering on phonons with the corresponding momentum relaxation time. More
recent examples are phonon-induced decay of the electron spin in quantum dots
\cite{Golovach,Sherman} or phonon limited mobility in a two-dimensional
electron gas (2DEG) with SOC \cite{Chen}. Among other examples
is the use of coherent acoustic phonons to create dynamic quantum dots in
SOC systems \cite{Stotz}.
\begin{figure}
\includegraphics[width=7.6 cm]{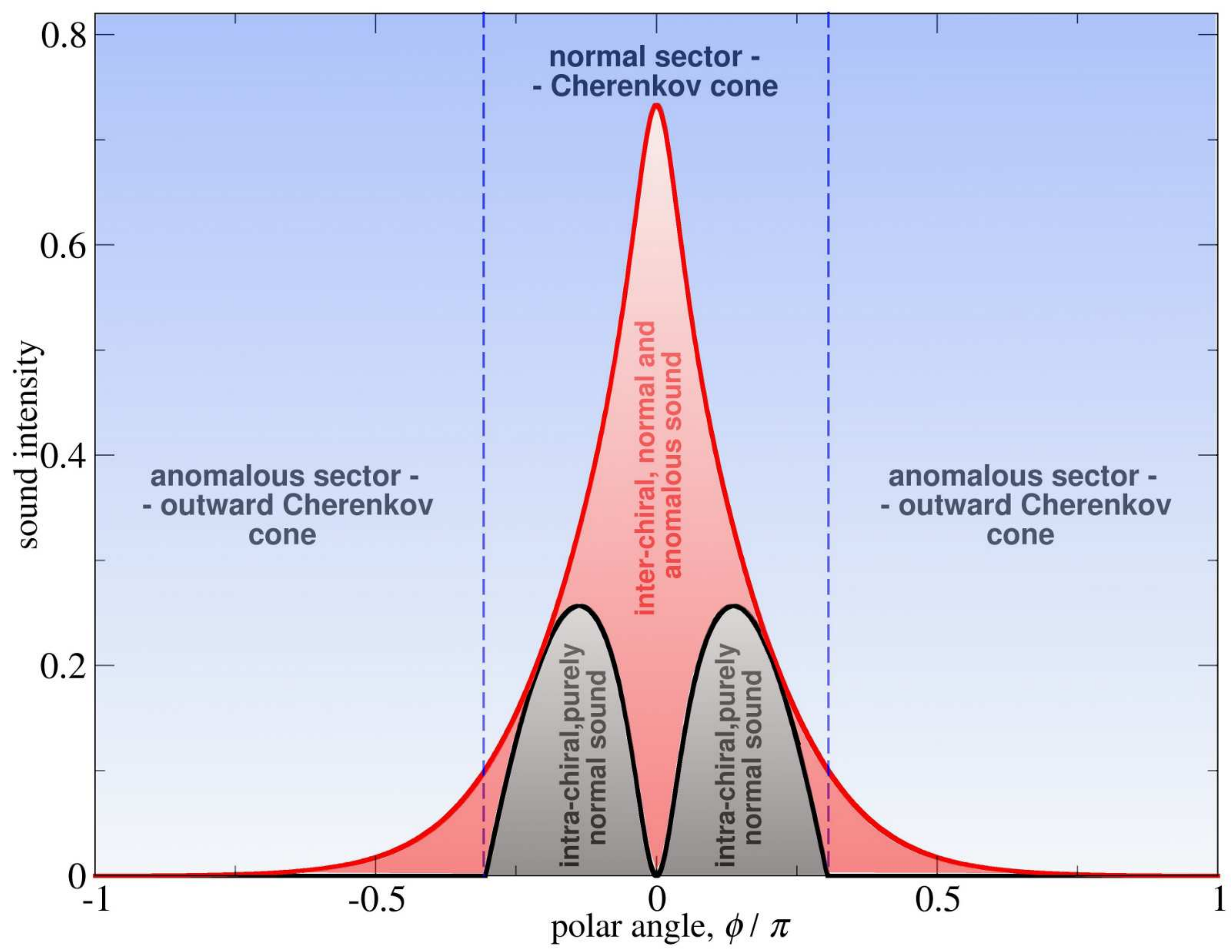}
\caption{\label{figure_2} (Color online) The intra- and inter-chiral
  contributions to ACE for $v=5\times 10^3$ m/s in a 2DEG with BRSOI in InAs
  structures with the following parameters: $c=4.2\times 10^3$ m/s,
  $\alpha=0.15\times 10^{-11} \text{eV m}$.}
\end{figure}

Here we take a different view angle on systems with SOI
and instead of electrons focus on phonons, in particular on the consequences
and fingerprints of SOI which could be observed in the
phonon dynamics.

There are obviously various aspects of the phonon dynamics in solids. In the
present investigation we will study one of them, an important and beautiful
phenomenon of the Cherenkov radiation.

Originally the Cherenkov effect was discovered by Cherenkov in the
electrodynamics of fast charged particles passing through transparent media
\cite{Cherenkov} and later theoretically explained by Tamm and Frank
\cite{Tamm}. It consists in the appearance of a forward light cone emitted by
a given medium under the impact of a charged particle moving with the velocity
larger than the speed of light in this medium. This Cherenkov effect, also
referred to as normal optical Cherenkov effect, is qualitatively different
from the well known deceleration radiation because in the latter case the
radiation is emitted by the particle itself \cite{Landau_VIII}.
\begin{figure}
\includegraphics[width=7.6 cm]{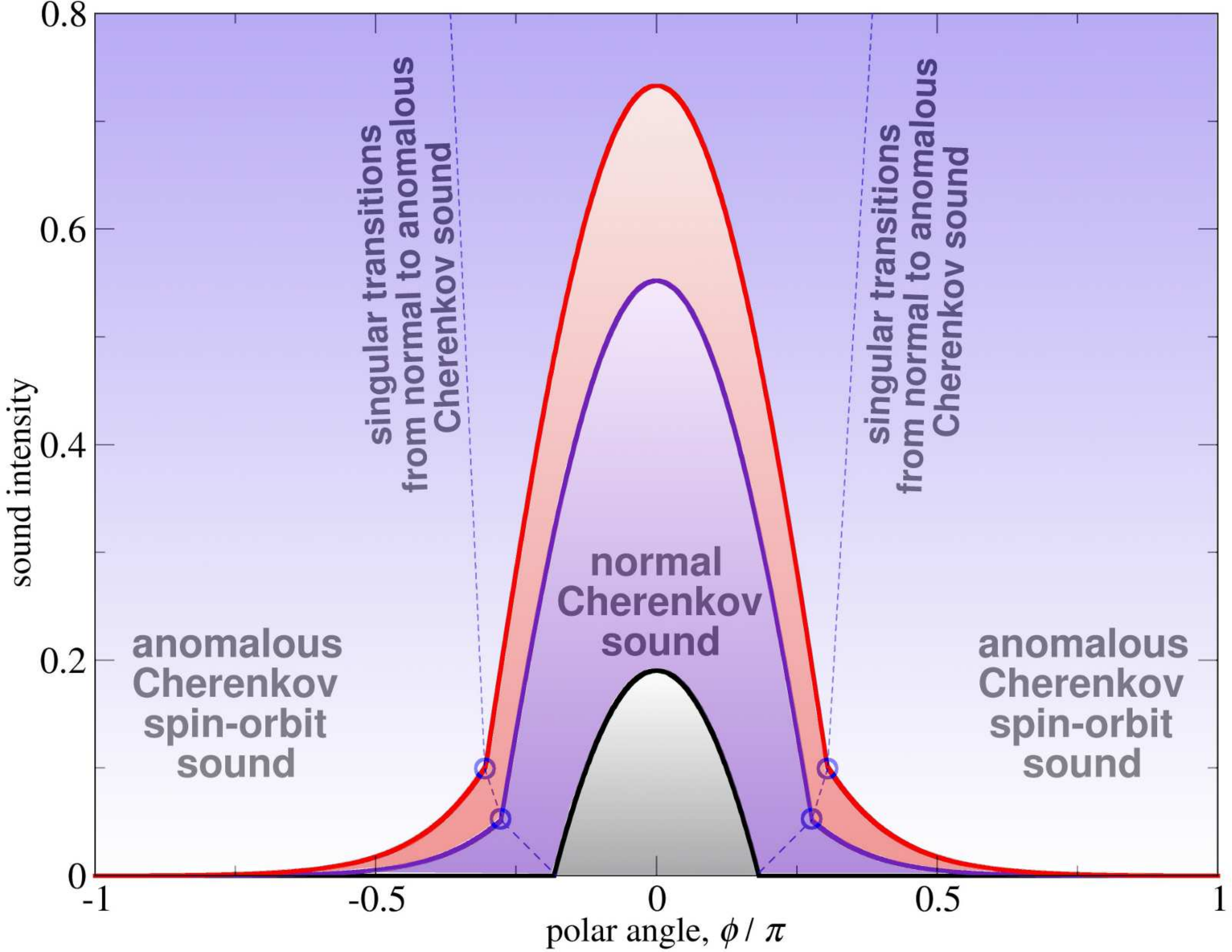}
\caption{\label{figure_3} (Color online) The total intensity of the Cherenkov sound for
  $v=5\times 10^3$ m/s in a 2DEG with BRSOI in InAs
  structures with $c=4.2\times 10^3$ m/s. The black (lower) curve corresponds to the
  absence of SOI, $\alpha=0.0\:\text{eV m}$. The lilac (intermediate)
  curve shows the situation for $\alpha=0.1\times 10^{-11} \text{eV m}$, while
  the red (upper) one is for $\alpha=0.15\times 10^{-11} \text{eV m}$.}
\end{figure}

With appearance of photonic crystals the optical Cherenkov effect was
rediscovered and an anomalous optical Cherenkov radiation with a backward
pointing cone was predicted \cite{Joannopoulos}. Here the anomalous radiation
is the result of strong inhomogeneity of the medium leading to complex Bragg
scattering.

It turns out that the normal Cherenkov effect has an acoustic counterpart. In
a three-dimensional (3D) medium an electron whose velocity larger than the
sound velocity of the medium excites a forward sound Cherenkov cone. The
sound intensity $I_s^{(3D)}$ is located inside a three-dimensional cone and
its azimuthal angular distribution \cite{LS} is
$I_s^{(3D)}(\theta)=\Theta(\theta_c-\theta)((v/c)\cos(\theta)-1)^2\sin(\theta)$,
where $0<\theta<\pi$ is the azimuthal angle, the angle between the incident
electron momentum $\vec{p}$ (chosen as the direction of the $z$-axes;
$\vec{p}=m\vec{v}$, $m$ is the particle mass and $|\vec{v}|=v$) and the phonon
momentum $\vec{q}$, $c$ is the sound velocity ($c<v$),
$\theta_c=\text{arccos}(c/v)$ is the Cherenkov cone angle and $\Theta(x)$ is
the Heaviside step function.

In two dimensions (2D) the sound intensity $I_s^{(2D)}$ is located inside a
two-dimensional Cherenkov cone with the polar angular distribution
$I_s^{(2D)}(\phi)=\Theta(\phi_c-|\phi|)((v/c)\cos(\phi)-1)$,
where $-\pi<\phi<\pi$ is the polar angle (here $\vec{p}$ is the direction of
the $x$-axis) and the Cherenkov angle $\phi_c$ of the two-dimensional
cone is given by the same expression as $\theta_c$ in three dimensions.

The essential difference between $I_s^{(3D)}$ and $I_s^{(2D)}$ is that in 2D
the strictly forward ($\phi=0$) sound emission is allowed while in 3D the
phonons with $\theta=0$ are dimensionally forbidden.

The characteristic feature of the normal acoustic Cherenkov effect (ACE) is
that both in the 3D and 2D cases the sound does not exist out of the Cherenkov
cone. However, as we demonstrate below, this situation radically changes as
soon as the spin and orbital degrees of freedom of the incident electron,
exciting the Cherenkov sound, are coupled by a certain SOI mechanism. In
this case it is for the first time shown that one
obtains an acoustic counterpart of the anomalous optical Cherenkov
radiation. However, what is remarkable, this anomalous ACE does not require
any inhomogeneity of the medium at all in contrast with the optical version of
the effect in photonic crystals \cite{Joannopoulos}.

To study the essential physics we consider a 2DEG with BRSOI. The
Bychkov-Rashba Hamiltonian is
$\hat{H}_0=\hat{\vec{p}}^2/2m+(\alpha/\hbar)[\hat{\vec{\sigma}}\times\hat{\vec{p}}]\cdot\vec{n}$.
Here $\vec{n}$ is the unit vector perpendicular to the 2DEG
plane, $m$ is the electron effective mass, $\hat{\vec{p}}$ is the momentum
operator, $\hat{\vec{\sigma}}$ is the Pauli matrix vector and
$\alpha\equiv\hbar p_\mathrm{so}/m$ is the SOC strength. In
the following we choose the direction of $\vec{n}$ to be the direction of the
$z$-axis. 

The Hamiltonian $\hat{H}_0$ lifts the twofold spin degeneracy
at momenta $\vec{p}\neq 0$ and results in a spin-orbit band splitting. The
spin is not a good quantum number anymore. It is well known that the chirality
operator,
$\hat{R}\equiv[\hat{\vec{\sigma}}\times\hat{\vec{e}}\;]\cdot\vec{n}$,
with $\hat{\vec{e}}\equiv\hat{\vec{p}}/|\hat{\vec{p}}|$ being the operator of
the momentum direction, commutes with the Hamiltonian $\hat{H}_0$ and
momentum operator. Its eigenvalues, $\lambda=\pm 1$, characterize the electron
energy spectrum,
$\epsilon_{\vec{p}\,\lambda}=\vec{p}^2/2m+\lambda p_\mathrm{so}|\vec{p}|/m$,
and the normalized eigenspinors of $\hat{H}_0$,
$\langle\vec{r}\sigma|\vec{p}\lambda\rangle=\exp(i\vec{p}\vec{r}/\hbar)(1/\sqrt{2})
[1,\,\lambda i\exp(-i\varphi_\vec{p})]$, where $\varphi_{\vec{p}}$ is the angle
between the electron momentum $\vec{p}$ and the $x$-axis and $\vec{r}$ is the
2D coordinate.

The physics of ACE comes from the electron coupling to acoustic phonons. The
phonon Hamiltonian has the standard second quantized expression \cite{Landau_V},
$\hat{H}_\text{ph}=\sum_{\vec{q}}\hbar\omega(\vec{q})(b^\dagger_\vec{q}b_\vec{q}+1/2)$,
where $b^\dagger_\vec{q}$, $b_\vec{q}$ are the phonon creation and annihilation
operators and the acoustic phonon spectrum is
$\hbar\omega(\vec{q})=c|\vec{q}|$ with $c$ being the sound velocity. The
Hamiltonian of the electron-phonon interaction reads \cite{AGD}
$\hat{H}_\text{el-ph}=g\sum_\sigma\int
d\vec{r}\hat{\psi}^\dagger_\sigma(\vec{r})\hat{\psi}_\sigma(\vec{r})\hat{\varphi}(\vec{r})$,
where $g$ is the electron-phonon coupling strength,
$\hat{\psi}^\dagger_\sigma(\vec{r})$, $\hat{\psi}_\sigma(\vec{r})$ are the
electronic field operators and
$\hat{\varphi}(\vec{r})=i\sum_{\vec{q}}\sqrt{\hbar\omega(\vec{q})/2V}[\exp(i\vec{q}\vec{r}/\hbar)b_\vec{q}-
\exp(-i\vec{q}\vec{r}/\hbar)b^\dagger_\vec{q}]$. 

Since ACE is an effect of the electron propagation through a medium the
natural mathematical language to describe this physical phenomenon is the
language of the Feynman propagators or the Green's functions. In the present
case the Green's functions containing physics of ACE are defined with respect
to physical vacuum \cite{LS}.

The corresponding self-energy diagram is shown in Fig. \ref{figure_1}. The
analytic expression corresponding to this diagram is obtained according to the
standard analytic reading rules \cite{AGD} and leads to the following
expression
\begin{equation}
\begin{split}
\Sigma_{\vec{p}\lambda}(\varepsilon)=&\frac{ig^2}{16\pi^3\hbar^2}
\sum_{\lambda'}\int\biggl[\frac{c^2\vec{q}^2}{\hbar^2\omega^2-c^2\vec{q}^2+i\eta}\times\\
&\times\frac{1+\lambda\lambda'\cos(\varphi_{\vec{p}-\vec{q}}-\varphi_\vec{p})}
{\varepsilon-\hbar\omega-\epsilon_{\vec{p}-\vec{q}\,\lambda'}+i\eta}\biggl]d\omega d\vec{q},
\end{split}
\label{self_energy}
\end{equation}
where $\eta$ is a positive infinitesimal. The frequency integration is
performed using the residue theorem. There is one pole in the lower half-plane,
$\omega_0=c|\vec{q}|/\hbar-i\tilde{\eta}$. Thus
\begin{equation}
\Sigma_{\vec{p}\lambda}(\varepsilon)\!=\!
\frac{g^2c}{16\pi^2\hbar^3}
\!\sum_{\lambda'}\!\!\int\!\frac{1\!+\!\lambda\lambda'\cos(\varphi_{\vec{p}-\vec{q}}-\varphi_\vec{p})}
{\varepsilon-c|\vec{q}|-\epsilon_{\vec{p}-\vec{q}\,\lambda'}+i\eta}|\vec{q}|d\vec{q}.
\label{self_energy_afi}
\end{equation}
The sound intensity is obtained from the imaginary part of
Eq. (\ref{self_energy_afi}) taken on the mass surface,
$\varepsilon=\epsilon_{\vec{p}\,\lambda}$, 
\begin{equation}
\begin{split}
&\text{Im}\Sigma_{\vec{p}\,\lambda}(\varepsilon=\epsilon_{\vec{p}\,\lambda})=-\frac{g^2c}{16\pi\hbar^3}\sum_{\lambda'}\int_0^{k_D}\int_0^{2\pi}\bigl[1+\\
&+\lambda\lambda'\cos(\varphi_{\vec{p}-\vec{q}}-\varphi_\vec{p})\bigl]\delta\bigl(E_{\vec{p}\,\lambda;\vec{q}\,\lambda'}\bigl)|\vec{q}|^2d|\vec{q}|d\phi,
\end{split}
\label{self_energy_im_p}
\end{equation}
where $k_D$ is the Debye momentum and
$E_{\vec{p}\,\lambda;\vec{q}\,\lambda'}\equiv\epsilon_{\vec{p}\lambda}-\epsilon_{\vec{p}-\vec{q}\,\lambda'}-c|\vec{q}|$. It
is enough to consider the Cherenkov sound excited only by an electron in the
chiral branch with $\lambda=+1$ since it contains all essential physics of ACE
while the Cherenkov sound excited by an electron with $\lambda=-1$ may be
obtained in a similar way and does not lead to new phenomena. Using the
expression $\delta\bigl[f(x)\bigl]=\sum_n(1/|f'(r_n)|)\delta(x-r_n)$,
where $r_n$ are the roots of the equation $f(x)=0$, we find
$\text{Im}\Sigma_{\vec{p}\,+}(\varepsilon=\epsilon_{\vec{p}\,+})=-(g^2m^2c^2/2\pi\hbar^3)\int_{-\pi}^\pi
I_s(\phi) d\phi$.
Here the dimensionless Cherenkov sound intensity,
$I_s(\phi)=I_s^{(\text{intra})}(\phi)+I_s^{(\text{inter})}(\phi)$, consists of
two contributions coming from intra- and inter-chiral electronic transitions,
\begin{equation}
\begin{split}
&I_s^{(\text{intra})}(\phi)=\Theta(\phi_c-|\phi|)\biggl[1+\\
&\frac{1-(c/v)q_1(\phi)\cos(\phi)}{\sqrt{1+(c/v)^2q_1^2(\phi)-2(c/v)q_1(\phi)\cos(\phi)}}\biggl]\frac{q_1^2(\phi)}{|h_1(\phi)|},
\end{split}
\label{I_intra}
\end{equation}
\begin{equation}
\begin{split}
&I_s^{(\text{inter})}(\phi)=\biggl[1-\\
&\frac{1-(c/v)q_2(\phi)\cos(\phi)}{\sqrt{1+(c/v)^2q_2^2(\phi)-2(c/v)q_2(\phi)\cos(\phi)}}\biggl]\frac{q_2^2(\phi)}{|h_2(\phi)|},
\end{split}
\label{I_inter}
\end{equation}
where $q_{1,2}(\phi)$ are the positive solutions of the equations
\begin{equation}
\begin{split}
&2\frac{v_\text{so}v}{c^2}+2\biggl(\frac{v}{c}\cos(\phi)-1\biggl)q_{1,2}-q^2_{1,2}\mp\\
&\mp2\frac{v_\text{so}v}{c^2}\sqrt{1+\bigg(\frac{c}{v}\biggl)^2q^2_{1,2}-2\cos(\phi)\frac{c}{v}q_{1,2}}=0,
\end{split}
\label{q_12}
\end{equation}
with $v_\text{so}\equiv p_\text{so}/m$ and $h_{1,2}(\phi)$ are the following
functions
\begin{equation}
\begin{split}
&h_{1,2}(\phi)=4\biggl[2\biggl(\frac{v}{c}\cos(\phi)-1\biggl)-2q_{1,2}(\phi)\mp\\
&\mp\!\frac{v_\text{so}v}{c^2}\frac{2q_{1,2}(\phi)(c/v)^2-2(c/v)\cos(\phi)}
{\sqrt{1\!+\!(c/v)^2q^2_{1,2}(\phi)\!-\!2\cos(\phi)(c/v)q_{1,2}(\phi)}}\biggl].
\end{split}
\label{h_12}
\end{equation}

The solution of Eqs. (\ref{q_12}) may be obtained numerically. It follows that
regardless of the existence of SOI the positive
solutions $q_1(\phi)$ are located inside the Cherenkov cone with an angle
$\phi_c$. This is reflected by the presence of the Heaviside step function in
Eq. (\ref{I_intra}). Therefore, the intra-chiral contribution contains only
the normal ACE. For $v_\text{so}\lesssim c$ the Cherenkov cone angle is 
$\phi_c=\arccos[c/(v+v_\text{so})]$ which assumes that the inequality 
$c<v+v_\text{so}$ must be satisfied. In the limiting case $v_\text{so}=0$ one
obtains from Eqs. (\ref{q_12}) $q_{1,2}=2[(v/c)\cos(\phi)-1]$. This leads to
the condition $(v/c)\cos(\phi)-1>0$ (requiring $v>c$). Therefore the
well-known result, $I_s^{(2D)}$, is reproduced.
\begin{figure}[h]
\includegraphics[width=7.6 cm]{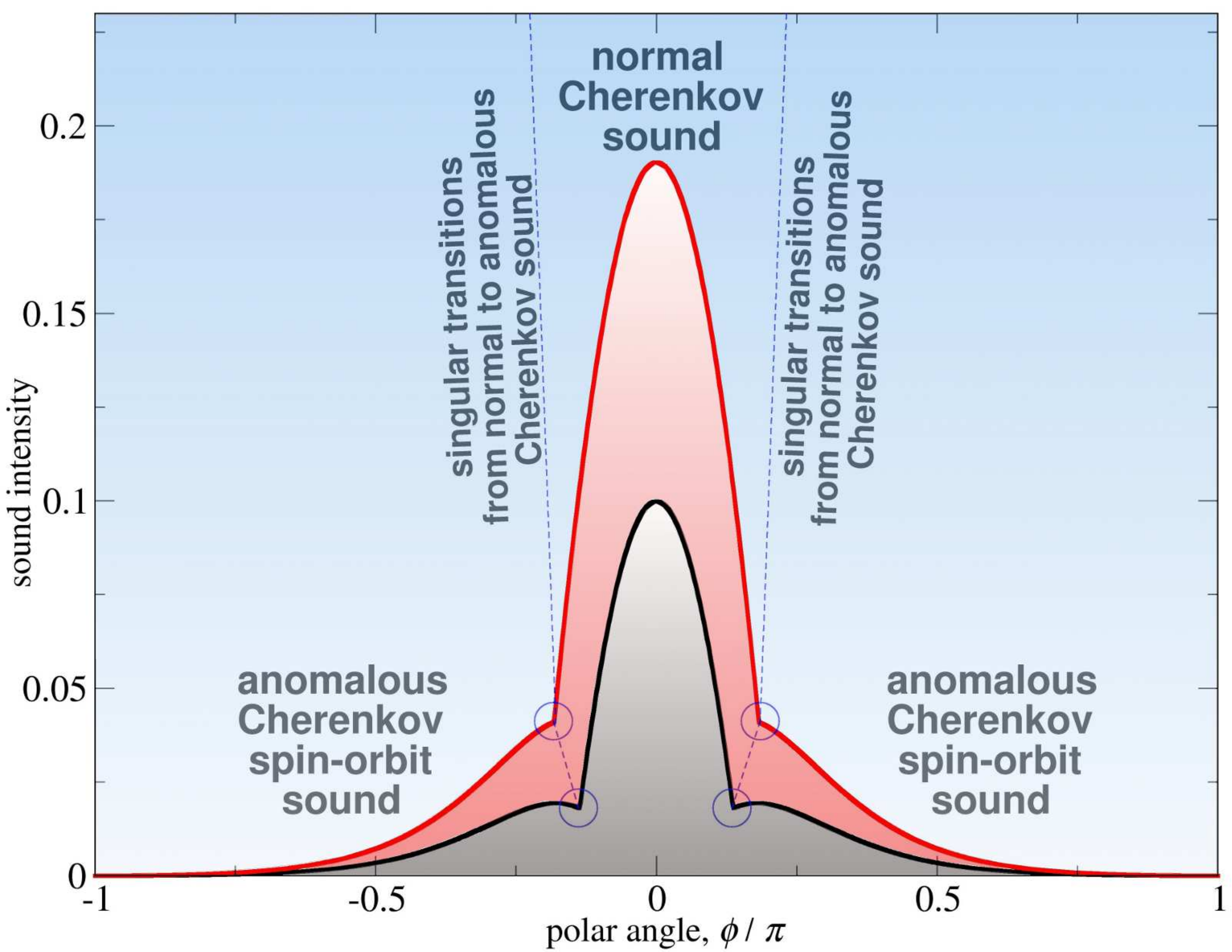}
\caption{\label{figure_4} (Color online) The same as in Fig. \ref{figure_3}
  but for $v=3.1\times 10^3$ m/s and for $\alpha_1=0.1\times 10^{-11} \text{eV
    m}$ (black or lower curve) and $\alpha_2=0.125\times 10^{-11} \text{eV m}$
  (red or upper curve).} 
\end{figure}

What is surprising is that for non-vanishing SOI
($v_\text{so}\neq 0$) the positive solutions $q_2(\phi)$ exist in the whole
interval $[-\pi,\pi]$. Therefore, the inter-chiral contribution to
the Cherenkov sound contains both normal (inside the Cherenkov cone) and
anomalous (outside the Cherenkov cone) ACE as is shown in Fig. \ref{figure_2}.

One can see from Eqs. (\ref{I_intra}) and (\ref{I_inter}) that because of the
Heaviside step function the first derivative of the total sound intensity
$I_s(\phi)$ has a singularity at the Cherenkov angle $\phi_c$.  This
singularity can be seen in Fig. \ref{figure_3} where $v>c$ and in a more
striking form in Fig. \ref{figure_4} where $v<c$.

\begin{figure}[h]
\includegraphics[width=7.6 cm]{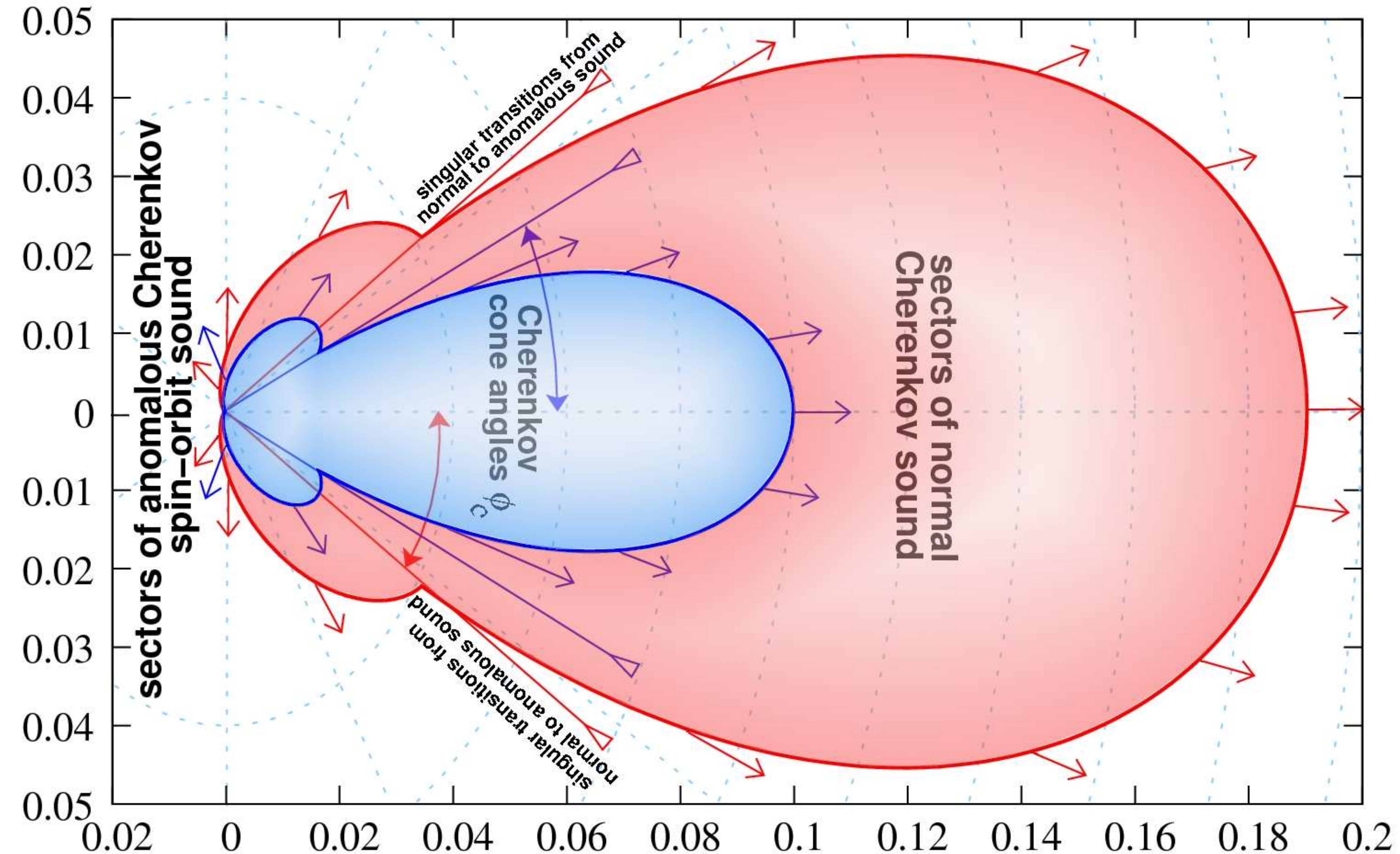}
\caption{\label{figure_5} (Color online) The curves from Fig. \ref{figure_4}
  in polar coordinates: blue (internal) for $\alpha_1$, red (external) for
  $\alpha_2$.}
\end{figure}

The two curves in Fig. \ref{figure_4} are plotted in Fig. \ref{figure_5} using
the polar coordinates in the 2DEG plane.

Clearly, the detection of those acoustic singularities is an experimental
challenge because it gives $\phi_c$ in a SOC system,
$\cos(\phi_c)=c/(v+v_\text{so})$, and thus the SOI strength.

In conclusion, the acoustic Cherenkov effect in a two-dimensional electron gas
with the Bychkov-Rashba spin-orbit interaction has been considered. It has
been shown that in this system in addition to the normal Cherenkov sound
inside the Cherenkov cone there also exists an anomalous Cherenkov sound
outside this cone. The singular transition from the normal to anomalous sound
at the Cherenkov angle provides an alternative experimental measurement of the
spin-orbit coupling strength. 

{\it Acknowledgments.} Support from the DFG under the program SFB 689 is
acknowledged.

\end{document}